\documentclass[prb,twocolumn]{revtex4}

\usepackage{graphicx}
\usepackage{amsmath}

\begin{document}

\title
{On solutions of the nonequilibrium x-ray edge problem}

\author
{Bernd Braunecker}

\affiliation
{Institute of Theoretical Physics, Faculty of Basic Sciences,
Swiss Federal Institute of Technology Lausanne, EPFL, CH-1015
Lausanne, Switzerland}

\date{\today}


\begin{abstract}
We rediscuss a nonequilibrium x-ray edge problem which in recent publications
led to discrepancies between the results of the perturbative and of an extended 
Nozi\`eres-De~Dominicis approach. We show that the this problem results
from an uncritical separation of momenta of the scattering potential, and we
propose a corrected Nozi\`eres-De~Dominicis solution.
\end{abstract}

\maketitle

\section{Introduction}

We address a question raised recently by Combescot and Roulet\cite{Combescot00} (CR)
in the context of the nonequilibrium x-ray edge problem.
The latter was studied some years ago by Ng\cite{Ng96} who developed
an extension of the well-known Nozi\`eres and De~Dominicis\cite{ND69} (ND) technique.
In the late 60s, ND devised this method 
for the computation of the x-ray absorption spectrum. They provided with it an elegant
description of the shape of the edge singularity in terms of the scattering
phase shift at the Fermi surface.
The approach is based on some subtle approximations, though, 
and thorough check of the condition of validity is required. 
Such a check is provided, for instance, by comparison
with the perturbative parquet diagram result of
Roulet, Gavoret and Nozi\`eres\cite{Roulet69,Nozieres69}, and
the validity of ND's solution for the original x-ray problem can 
in this sense be proven.
For the nonequilibrium case, however, CR revealed an inconsistency by
comparing a similar parquet diagram summation to Ng's result, yet they
did not provide an explanation of its origin.
In the present paper, we show that the discrepancies arise from an 
uncritical use of the decoupling of the momenta of the scattering potential.
Further, we show how ND's method can be modified to yield a result
coinciding with the perturbative one. 
This is illustrated on CR's academic problem for which it turns out that 
Ng's multi-channel extensions are not required\footnote{%
In our opinion, there remains an inconsistency with Ng's extensions,
independently to the topic discussed here, which arises from
a too specialized Ansatz to the solution of the problem.
A detailed discussion of this discrepancy is provided
in a forthcoming publication. The conceptual difficulty addressed 
in the present paper, however, persists also in a modified Ansatz.
}%
.

\section{Nonequilibrium x-ray problem}

\begin{figure}[p]
	\includegraphics{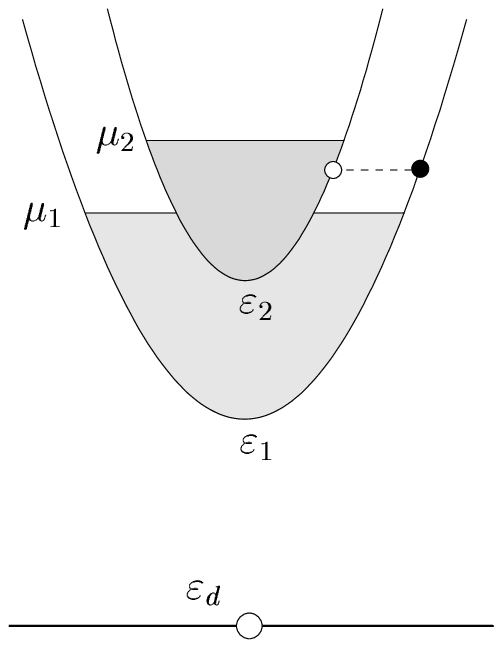}
	\caption{Two nonequilibrium Fermi seas extending between the energies $(\varepsilon_1,\mu_1)$
	and $(\varepsilon_2,\mu_2)$, respectively.
	A zero energy particle-hole pair between the subbands is indicated. Pairs with
	positive or negative energies are possible.
	\label{fig:system}}
\end{figure}

The nonequilibrium system to be discussed consists of two Fermi seas $n = 1,2$,
referred to as \emph{subbands} which are characterized
by the fixed chemical potentials $\mu_1<\mu_2$ and the energies $\varepsilon_1, \varepsilon_2$
at the bottom of the subbands.
We assume that $\mu_1 > \varepsilon_2$, so there is a nonzero overlap between the bottom 
of the subband 2 and the Fermi surface $\mu_1$.
A core state with energy $\varepsilon_d$ and an infinite mass is assumed to exist below the subbands.
The qualitative situation is shown in Fig.~\ref{fig:system}.

If both subbands were isolated, the x-ray absorption spectrum would be
the result of a simple superposition of two independent x-ray problems, and
two edge singularities corresponding to $\mu_1$ and $\mu_2$ could be observed.
The mixture between the two subbands, however, allows for a new physical
process:
Particle-hole excitations with the particle in one subband and the hole
in the other one become possible with energies that may be positive or
negative (see Fig.~\ref{fig:system}). 
The form of the absorption spectrum is changed considerably. The edge 
singularities are broadened, the  core-hole acquires a finite life-time,
and absorption below the threshold becomes possible, where the
missing energy is compensated by such particle-hole excitations of negative
energy. 

The Hamiltonian of the system is assumed to be given by 
\begin{equation} \label{eq:Hamiltonian}
	H = \sum_{k n} \varepsilon_{k n} \, c_{k n}^\dagger c_{k n} 
	+ \varepsilon_d \, b^\dagger b 
	+ \sum_{k k' n n'} V^{n n'}_{k k'} \,
	c_{k n}^\dagger c_{k' n'} b b^\dagger,
\end{equation}
where $n = 1, 2$ labels the subbands. Conduction electrons with momenta $k$ 
in the subbands $n$
are created and annihilated by $c_{k n}^\dagger$ and $c_{k n}$, 
while $b^\dagger$ and $b$ operate on the core state.
The interaction with the photon field is described by the operator
\begin{equation} \label{eq:xray_field}
	H_x = \sum_{k n} \lambda_k^n \, c_{k n} b^\dagger + \text{h.c.}
\end{equation}
For simplicity, spin indices have been dropped.

\section{Perturbative approach of Combescot and Roulet}
\label{sec:pert}

The problem, as it is defined by the Hamiltonians \eqref{eq:Hamiltonian}
and \eqref{eq:xray_field} above, depends on the parameters
$\lambda_k^1, \lambda_k^2, V^{11}_{k k'}, V^{22}_{k k'}$ and $V^{12}_{k k'}$. With respect to 
the original x-ray problem the new physics must enter
with the subband couplings $V^{12}_{k k'}$. To focus on the new effects we 
consider with CR the academic problem of $V^{11}_{k k'} = V^{22}_{k k'} = \lambda^{2}_{k} = 0$.
This means that a photon can excite the core electron only into the subband $n = 1$,
and particle-hole excitations are due only to particles tunneling between both subbands,
as indicated in Fig.~\ref{fig:system}.

The perturbative approach of CR consists in assuming $V^{12}_{k k'} \equiv -V^{12}$ and in
expanding the response function $\mathcal{S}(\Omega)$, which is the Fourier transform
of
\begin{equation}
	\mathcal{S}(t_2-t_1) = \left\langle T\{ H_x(t_2) H_x(t_1)\}\right\rangle,
\end{equation} 
in powers of 
\begin{equation}
	\tilde{V} = V^{12} V^{21} = |V^{12}|^2.
\end{equation}
The imaginary part of $\mathcal{S}(\Omega)$ yields the absorption rate $\mathcal{A}(\Omega)$.
CR performed the sum over the most singular parquet diagrams and found an
absorption rate of the x-ray problem which is, close to the threshold energy 
$\omega_0 = \mu_1 - \varepsilon_d$, of the form
\begin{equation} \label{eq:A_pert}
	\mathcal{A}(\Omega)
	\approx \mathrm{Im} \frac{\nu_{11}}{2g} \left(\frac{\xi_0}{|\Omega-\omega_0|}\right)^{2g}
\end{equation}
for $\xi_0$ a cutoff of the order of the band width,
$\theta$ the usual step function, 
and $g$ the effective coupling constant
\begin{equation} \label{eq:g}
	g = \nu_{11} \tilde{V} (\bar{\nu}_{21} - i \pi \nu_{21}).
\end{equation}
Here we have used the notation $\nu_{n n'} = \nu_n(\mu_n')$ for 
the density of states in the subband $n$ at the energy $\mu_{n'}$, and 
$\bar{\nu}_{n n'}$ for the integral
\begin{equation} \label{eq:bar_nu}
	\bar{\nu}_{n n'}
	= \int_{\varepsilon_{n}}^\infty \mathrm{d}\varepsilon \, \nu_n(\varepsilon)\,
	P \frac{1}%
	     {\mu_{n'} - \varepsilon},
\end{equation}
with $P$ denoting the principal value.
The approximation \eqref{eq:A_pert} holds for small values of the coupling constant 
$g$. 

CR compared this result with Ng's extended ND approach.
Their criticism refers to the appearance of
the density of states $\nu_{22} = \nu_2(\mu_2)$, and not, as in the perturbative
result above, of the density of states at $\mu_1$, $\nu_{21} = \nu_2(\mu_1)$. 
The presence of $\nu_{22}$ implies a coupling
of particles at the respective energies $\mu_1$ and $\mu_2$, whereas the zero energy
inter-subband fluctuations at $\mu_1$ are entirely absent. This is definitely 
unphysical as soon as $\mu_2-\mu_1$ becomes large.
The origin of this discrepancy remained unresolved in CR's paper
and is discussed in the next section.

\section{Origin of the unphysical behavior}

The problems with ND's approach arise from an uncritical use of the momentum
decoupling 
\begin{equation} \label{eq:decouple}
	V^{n n'}_{k k'} \to V^{n n'} u^n_k u^{n'}_{k'}.
\end{equation}
As CR have noticed with the perturbative results above, the important particle-hole excitations
occur in the low energy sector with the particle and hole energies close to the Fermi 
surface $\mu_1$. In the single-band equilibrium x-ray problem, the restriction to this 
energy sector can be assured by the decoupling \eqref{eq:decouple}.
As it turns out, the explicit form of the $u_k$ is of little importance
because the form of the free propagator is controlled by the discontinuity at the Fermi 
surface (see Ref.~\onlinecite{ND69}, Eq. \mbox{(33)} and below) 
which naturally imposes the low-energy restrictions.
There is consequently no risk in assuming the potential $V_{kk'}$ to be constant in the whole
band as long as one introduces the adequate cutoff $\xi_0$ imposed by the band width. 

In the nonequilibrium case a similar ``naive'' splitting of the form
\eqref{eq:decouple} leads to the
unphysical results mentioned above. If one assumes the potential to be constant over the
whole band widths, the free propagators which are integrated over the $u_k^n$ 
are controlled by the discontinuities at the Fermi surfaces 
of their respective subbands analogously to the 
original x-ray problem. The coupling of the subbands then leads to the mixture
of the energy sectors close to the different Fermi surfaces, and therefore to large
energy particle-hole excitations of the order of $\mu_2-\mu_1$, whereas the
inter-subband fluctuations close to $\mu_1$ and $\mu_2$ respectively
are absent. From CR's perturbative approach, however, we know that 
the latter are the relevant processes and cannot be neglected.

In the sequel we propose a projection onto these low-energy sectors which 
is artificial in a similar way as ND's splitting \eqref{eq:decouple}
in the one-band case but allows us to capture the important particle-hole 
excitations found within the perturbative approach.

\section{Modification of the Nozi\`eres-De~Dominicis approach}
\subsection{Splitting of the potential. Asymptotic forms of propagators}

Let us reconsider the academic problem above in which only the potentials
$V^{12}_{k k'} = [V^{21}_{k' k}]^*$ are non-zero.
To restrict to energies close to $\mu_1$ we assume these potentials to be
separable as
\begin{equation}
	V^{12}_{k k'} = -V^{12} \, u^{11}_k u^{21}_{k'},
\end{equation}
where the $u^{n n'}_k$ are functions which refer to values of $k$ in the 
subband $n$ and which are maximum at the Fermi surface $\mu_{n'}$
and fall off sufficiently fast away from it.
To maintain the hermiticity of the potentials $V^{12}$ we assume these functions
to be real and equal, $u^{11}_k = u^{21}_k \equiv u^{1}_k$.

In ND's approach, the constant part $V^{12}$ acts as an ``external'' potential
on particles described by the operators $c_n^\dagger = \sum_k u^1_k c_{k n}^\dagger$
during times between $t_1$ and $t_2$. The relevant physical quantities 
can be obtained from the so-called transient propagator, $\varphi(t,t')$, which describes the 
time evolution of these particles, with $t_1 < t,t' < t_2$.
The potential $V^{12}$ enters as the self-energy part into the Dyson equation for 
the transient propagator (see Eq.~\eqref{eq:dyson} below). 
The free part of this Dyson equation is described by 
the Green's functions
\begin{multline} \label{eq:def_G}
	G_{n}(t) 
	= \sum_k (u^1_k)^2 \ G_{k n}(t) 
	= - i \sum_k (u^1_k)^2 \ \mathrm{e}^{-i \varepsilon_{k n} t} \ \\
	\times
	\Bigl[ 
		\theta(t)  \theta(\varepsilon_{k n} - \mu_n) -
		\theta(-t) \theta(\mu_n - \varepsilon_{k n})
	\Bigr].
\end{multline}
Following ND we must examine the asymptotic behavior of these functions
as $t \to \infty$.
We first consider the case $n=1$. Here the cutoff function $u^1_k$ is
centered at the Fermi surface $\mu_1$ which corresponds to the situation
of the original x-ray problem. Let us recall ND's arguments to obtain
an approximate expression for $G_1(t)$: For large times $t$ the
behavior of $G_1(t)$ is dominated by the discontinuity at the Fermi surface.
Let $\varrho_{11}(\varepsilon) = \nu^1(\varepsilon) (u^1(\varepsilon))^2$ with
$\nu^1(\varepsilon)$ the density of states in the subband 1, and 
$u^1(\varepsilon_{k1}) = u^1_k$. Then, for $t>0$
\begin{equation} \label{eq:G_1_int}
	G_1(t) = -i \int_{\mu_1}^\infty \mathrm{d}\varepsilon \, \varrho_{11}(\varepsilon)
	\mathrm{e}^{-i\varepsilon t},
\end{equation}
which becomes for large times $t \gg \xi_0^{-1}$ (where $\xi_0$ is the
characteristic energy describing the decay of $u^1(\varepsilon)$),
\begin{equation} \label{eq:G_1_asymp}
	G_1(t) \approx \frac{\nu^1(\mu_1)}{t} \, \mathrm{e}^{-i \mu_1 t}.
\end{equation}
The same expression holds for $t<0$. 
We observe that the exact shape of $u^1(\varepsilon)$ is of no importance
as long as $|t| \gg \xi^{-1}$.
To include the short-time behavior, $t \lesssim \xi_0^{-1}$, 
ND imposed that integrals over the product of
$G_1(t)$ and some slowly varying function (the transient propagator) must yield
the correct result. Hence, if $a$ is a time cutoff such that $a \gg \xi_0^{-1}$
but still much smaller than the characteristic time scale of the transient
propagator, the short-time behavior of $G_1(t)$ enters only through the integral
\begin{equation}
 	A_1 
	= \int_{-a}^a \mathrm{d}t \, G_1(t)
	= \int_{-\infty}^\infty \mathrm{d}\varepsilon \, 
	\varrho_{11}(\varepsilon) \, P\frac{1}{\varepsilon-\mu_1}
	\equiv - \bar{\nu}_{11},
\end{equation}	
where we have used Eq.~\eqref{eq:G_1_int} for the second equality.
This quantity can be added to the asymptotic expression \eqref{eq:G_1_asymp}
in the form of $A_1 \delta(t)$,
and we obtain the central approximation of ND's approach
\begin{equation} \label{eq:G_1}
	G_1(t) = - \nu_{11} \, P\frac{1}{t} \, \mathrm{e}^{-i \mu_1 t} - \bar{\nu}_{11} \, \delta(t).
\end{equation}

The case $n=2$ is more subtle because there is no Fermi edge discontinuity
at $\mu_1$ for electrons in the subband $2$. Let us set
$\varrho_{21}(\varepsilon) = \nu_2(\varepsilon) (u^1(\varepsilon))^2$ with $\nu_2(\varepsilon)$
the density of states in the subband 2.
We have, from Eq.~\eqref{eq:def_G},
\begin{equation}
\begin{aligned}
	G_2(t) = \ &-i \theta(t)  \int_{\mu_2}^\infty    \mathrm{d}\varepsilon \, 
					\varrho_{21}(\varepsilon) \, \mathrm{e}^{-i \varepsilon t}\\
	         \ &+i \theta(-t) \int_{-\infty}^{\mu_2} \mathrm{d}\varepsilon \, 
			 		\varrho_{21}(\varepsilon) \, \mathrm{e}^{-i \varepsilon t}.
\end{aligned}
\end{equation}
If $\mu_2-\mu_1$ is large with respect to the decay of the function $u^1(\varepsilon)$ the 
first integral can be neglected. Using the same argument, we can push
the upper boundary of the remaining integral to infinity. Hence
\begin{equation} \label{eq:G_2_a}
	G_2(t) = +i \theta(-t) \int_{-\infty}^{\infty} \mathrm{d}\varepsilon \, \nu_2(\varepsilon) (u^1(\varepsilon))^2 \, \mathrm{e}^{-i \varepsilon t},
\end{equation}
For large times, $|t| \gg \xi_0^{-1}$, the period of oscillation of the exponential
is much shorter than the time of variation
of the $u^1(t)$ so that the integral averages to zero. 
Following the above argumentation, the short-time behavior of $G_2(t)$
can be resumed into a term $A_2 \delta(t)$ with (using Eq.~\eqref{eq:G_2_a})
\begin{equation} \label{eq:A}
	A_2 = \int_{-a}^0 \mathrm{d}t \, G_2(t)
	= i \int_{-a}^0 \mathrm{d}t \int_{-\infty}^\infty \mathrm{d}\varepsilon \, 
	\varrho_{21}(\varepsilon) \mathrm{e}^{-i t (\varepsilon-\mu_1)}.
\end{equation}
The exchange of the two integrals demands some care because the integrand becomes
singular. For nonzero values of $\varepsilon$ the time integration can be performed first
and yields an integrand proportional to $1/\varepsilon$. In order to handle $\varepsilon\to 0$, 
we notice that only the advanced part of the function 
survives because of the constraint $t<0$. 
This implies that the energies entering in the exponential
of Eq.~\eqref{eq:A} must be shifted by an infinitesimal imaginary 
amount, $\varepsilon \to \varepsilon + i \eta$, with $\eta >0$. 
The time integration can then be performed first, and we obtain
\begin{equation} 
\begin{aligned}
	A_2 
	&= - \int_{-\infty}^\infty \mathrm{d}\varepsilon \, 
	\varrho_{21}(\varepsilon) 
	\frac{1}{\varepsilon-\mu_1+i \eta},
\end{aligned}	
\end{equation}
where we have neglected the term at the lower boundary of the time integral since
it is due to the sharp artificial time-cutoff $a$. With the 
formula $1/(x-i\eta) = P(1/x) + i \pi \delta(x)$ we finally find
\begin{equation}
	A_2 = - \bar{\nu}_{21} + i \pi \nu_{21}
\end{equation}
with 
\begin{equation}
	\bar{\nu}_{21} =  \int_{-\infty}^\infty \mathrm{d}\varepsilon \, \varrho_{21}(\varepsilon)
	P\frac{1}{\varepsilon-\mu_1},
\end{equation}
which coincides with $\bar{\nu}_{21}$, given by Eq.~\eqref{eq:bar_nu}
with a slightly modified definition because of the weight function $\varrho_{21}(\varepsilon)$.

The approximate form resulting from ND's argumentation is therefore given by
\begin{equation} \label{eq:G_2}
	G_2(t) = \left[i \pi \nu_{21} - \bar{\nu}_{21} \right] \, \delta(t).
\end{equation}
There is no algebraic decay in $1/t$ because the Fermi surface discontinuity 
for the subband $n =2$ is of no relevance for processes of energies close to
$\varepsilon = \mu_1$. 

\subsection{Transient propagators and absorption rates}

With the approximate forms of the initial Green's functions,
Eqs.~\eqref{eq:G_1} and \eqref{eq:G_2}, we can follow ND's analysis
step by step, and we focus here only on the new features in the derivation.

The computation of physical quantities is based on the transient propagators 
$\varphi^{n n'}$ which can be defined by the Dyson equations
\begin{multline} \label{eq:dyson}
	\varphi^{n n'}(t,t') = G_{n}(t-t') \delta_{n n'}\\
	- \sum_{n''} \int_{t_1}^{t_2} \mathrm{d}t'' \, G_{n}(t-t'') \, V^{n n''} \, 
	\varphi^{n'' n'}(t'',t').
\end{multline}
For the academic problem under discussion, the functions $\varphi^{11}$ and $\varphi^{21}$
are found to form a closed set of equations,
\begin{equation}
\begin{aligned}
	\varphi^{11}(t,t') &= G_1(t-t') - \int_{t_1}^{t_2} \mathrm{d}t'' \,
	G_1(t-t'') \, V^{12} \, \varphi^{21}(t'',t'),\\
	\varphi^{21}(t,t') &= - \int_{t_1}^{t_2} \mathrm{d}t'' \,
	G_2(t-t'') \, V^{21} \, \varphi^{11}(t'',t').
\end{aligned}
\end{equation}
Since $G_2(t)$ is proportional to a $\delta$-function, we obtain a single integral equation
for $\varphi^{11}$, 
\begin{equation}
	\varphi^{11}(t,t') = G_1(t-t') - \int_{t_1}^{t_2} \mathrm{d}t'' \,
	G_1(t-t'') \, \tilde{V}_{11} \, \varphi^{11}(t'',t'),
\end{equation}
in which the effective interaction
\begin{equation}
	- \tilde{V}_{11} = (-V^{12}) [i \pi \nu_{21} - \bar{\nu}_{21} ] (-V^{21})
\end{equation}
leads to the same coupling constant $g$ as in the perturbative approach, given by Eq.~\eqref{eq:g}.
The remaining problem coincides exactly with ND's original x-ray problem
with the interaction potential $V$ replaced by the 
effective potential $\tilde{V}_{11}$. Therefore, the absorption rate becomes,
from Eq.~\mbox{(66)} of Ref.~\onlinecite{ND69},
\begin{equation}
	\mathcal{A}(\Omega) \sim \mathrm{Im} \left[
	\left(\frac{\xi_0}{|\Omega-\omega_0|}\right)^{\epsilon}
	\mathrm{e}^{i \pi \epsilon \theta(\Omega-\omega_0)} \right],
\end{equation}
with 
\begin{equation}
	\epsilon = 2 \frac{\delta}{\pi} - \frac{\delta^2}{\pi^2},
\end{equation}
and $\delta$ the complex ``phase shift'' at $\varepsilon = \mu_1$ induced by the effective
potential $\tilde{V}_{11}$,
\begin{equation}
	\frac{\tan\delta}{\pi} = \frac{\nu_{11} \tilde{V}_{11}}{1+\bar{\nu}_{11} \tilde{V}_{11}}.
\end{equation}
For small $V^{12}$, one has $\delta/\pi \approx \nu_{11} \tilde{V}_{11}$, and
the absorption rate merges with the parquet graph result of CR.
Furthermore, the imaginary part of $\delta$ leads to a broadening of the
sharp Fermi edge singularities which exists in the case of disconnected subbands.

\section{Conclusion}

With the preceding discussion of the nonequilibrium x-ray problem we have seen
that ND's technique provides an elegant calculation tool for which, however, the 
conditions of validity must be checked carefully at the beginning. 

The key approximation is the restriction to 
low-energy fluctuations close to the Fermi surface(s). The separation of momenta
in the interaction potentials must occur in these energy sectors.
In the single-band case of the original x-ray problem, this
low-energy restriction enters naturally by the discontinuity of the Green's
functions at the Fermi surface which controls the large time behavior
of these functions. The particular form of the weight functions $u^n(\varepsilon)$
which project onto the low-energy sector has only little importance, so that it even might
be taken as constant over the whole band, as often assumed in the literature.

We have noticed with CR that the latter assumption leads to unphysical results within ND's
approach in the case of nonequilibrium systems. Here the form of the
weight functions becomes important, and we have shown how the approximations
of the initial Green's functions must be chosen to obtain physical results. 
In particular, the 
decay in $1/t$ which is characteristic for ND's asymptotic approximation,
appears only if a Fermi surface lies
within the energy range in which the cutoff functions $u^n(\varepsilon)$ are non-zero.

With this taken into account, ND's method remains valid and provides
us with quantitative results for the exponents of the power-law divergences
at the Fermi edges.
Moreover, the absence of the $1/t$-decay in the subband $n=2$
reduces the complexity of the discussed multi-channel problem. 
Instead of systems of singular integral equations, we deal with
a scalar (single-channel) problem in which the presence
of the other channels is reflected by a mere renormalization of the 
interaction strength which, however, becomes a complex quantity.

\begin{acknowledgments}
I thank P.-A. Bar\`es for many helpful discussions.
The support of the Swiss National Fonds is gratefully acknowledged.
\end{acknowledgments}


\end{document}